\documentclass[12pt,a4paper]{article}
\usepackage[pdftex]{graphicx,color}     
\graphicspath{{fig/}}               

\begin{document}

\title{On the conflict between precision and robustness\\ in the
  proportion regulation of cell types}
\author{Ismael R\`afols$^1$\thanks{rafols@sawada.riec.tohoku.ac.jp}, Harry K. MacWilliams$^2$,\\ Yasuo Maeda$^3$ and Yasuji Sawada$^1$\\ \\
  \and $^1$Research Institute of Electrical Communication, Tohoku University,\\
  2-1-1 Katahira, Aoba-ku, Sendai 980-8577, Japan
  \and $^2$Zoologisches Institut, Ludwig-Maximilians-Universit\"at,\\
  Luisenstra\ss e 14, 80333 M\"unchen, Germany
  \and $^3$Biological Institute, Graduate School of Science, Tohoku University, \\
  Aoba, Sendai 980-8578, Japan}

\date{April, 2001} \maketitle

\newpage
\begin{abstract}
  In {\it Dictyostelium discoideum} the proportion of cell types is
  known to be actively regulated, recovering, for example, after the
  removal of a given cell type.  However, we have recently shown that
  regulation is intrinsically imprecise: it controls the proportion
  above/below certain upper and lower thresholds, but not within an
  allowed range of values that these thresholds define. To explain
  this finding we present a model based on (i) a global negative
  feedback, and (ii) a cell-autonomous positive feedback leading to a
  hysteresis-like behaviour (i.e, a bistability of cell type). The
  simulated cell-type proportion is indeed found to span a range of
  values as a consequence of the bistability. We conclude that there
  is a general conflict between the precision of proportion regulation
  and the robustness of the differentiation of cell types.
\end{abstract}

\newpage Tissues of multicellular organisms are formed by a variety of
cell types in fixed proportions. Both the normal maintenance of these
tissues and their restoration after injury involve
transdifferentiation, dedifferentiation and/or differentiation of
pluripotent cells (the so-called stem cells) into finally
differentiated cell types \cite{snc00}. Examples of major biomedical
relevance include the blood cell system and the epidermis. While the
molecular details of the underlying cell-cell signalling are now being
unveiled \cite{frm00}, the global mechanisms that regulate the
populations remain obscure. Some theoretical investigations on
proportion regulation have been conducted \cite{sch96,mzg95}, but
their scope has often been too abstract/complex for comparison with
experimental data or otherwise too detailed to give conceptual
insights. Here we present a simple model, based on recent experimental
findings for the regulation of cell types in the slime mould {\it
  Dictyostelium discoideum}.

\begin{figure}[htbp]
\begin{center}
  \includegraphics[width=14cm]{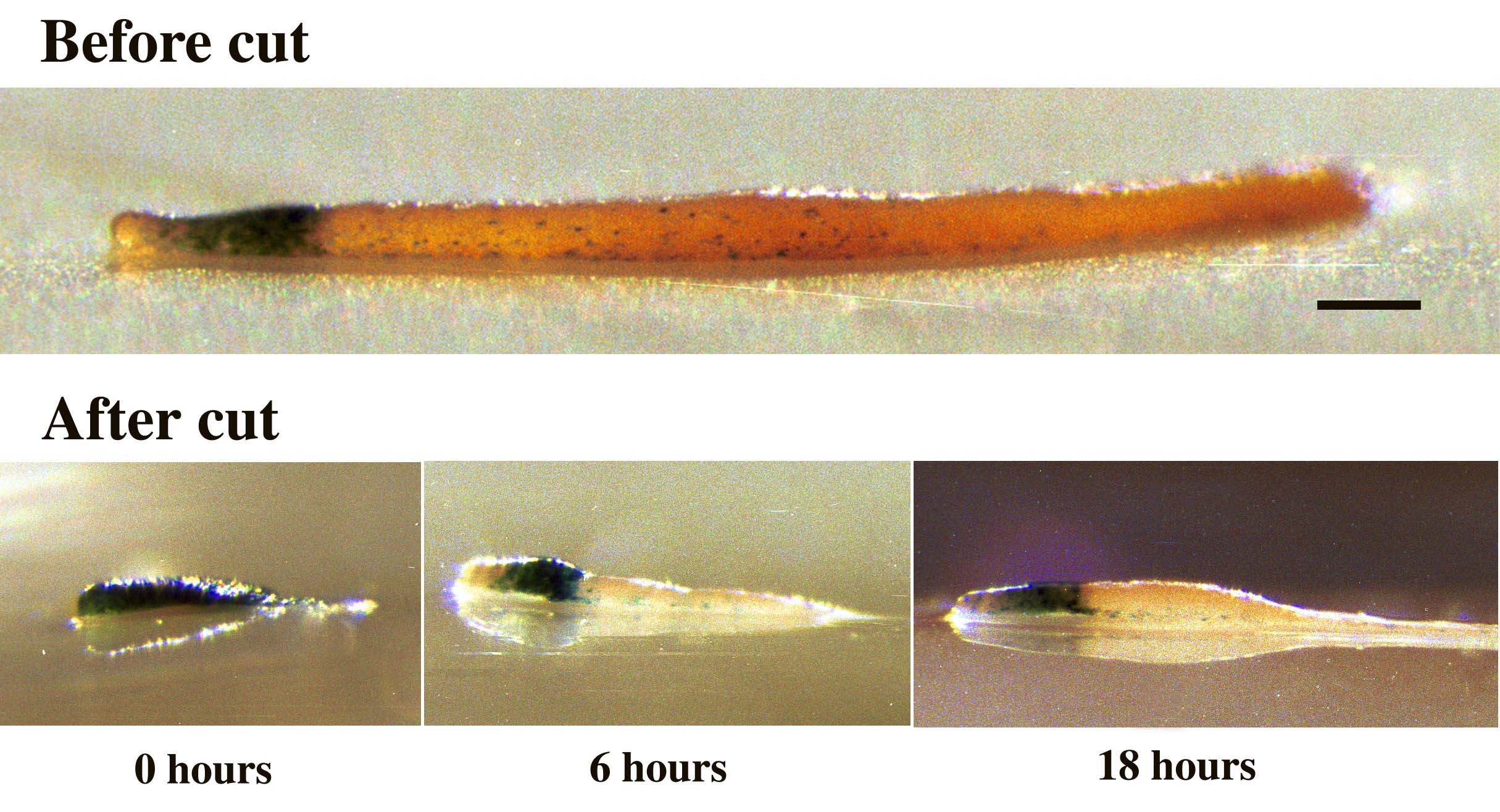}
  \includegraphics[width=13cm]{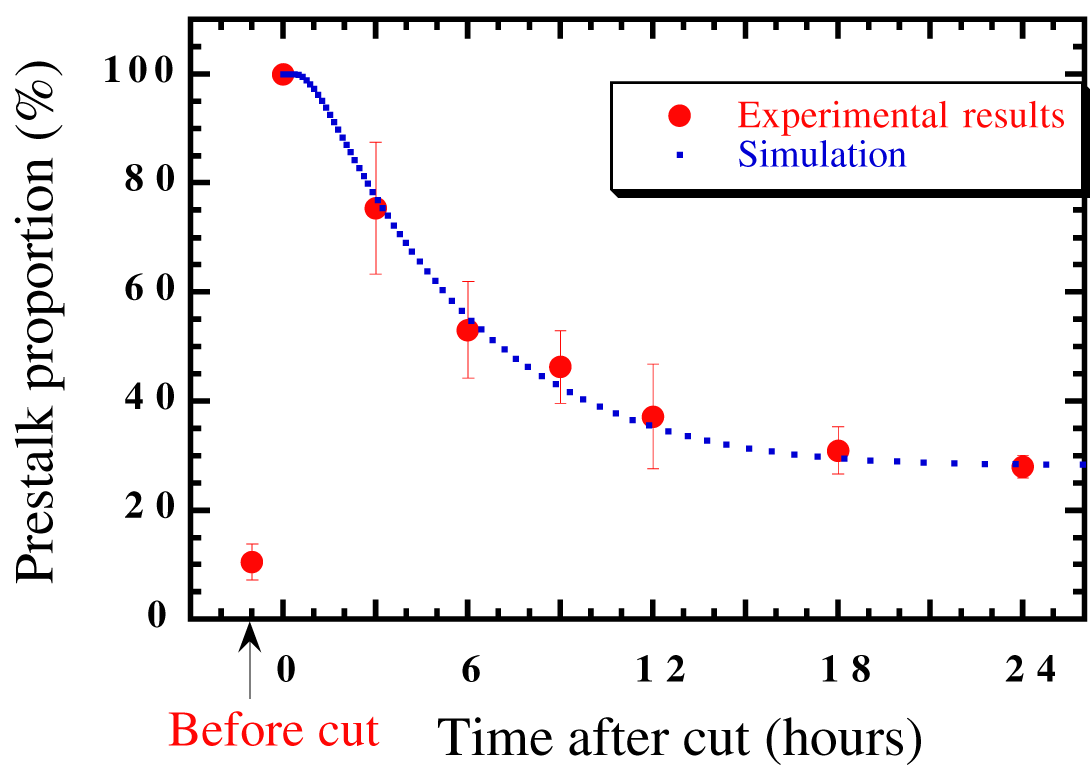}
\caption{ (Top panel) Slugs of {\it Dictyostelium discoideum} before
  and during regeneration after total removal of the prespore cells.
  (Bottom panel) Prestalk proportion before and during regeneration.
  Small squares show the results of simulation ($\tau=30$, $u_1=2.5$,
  $u_2=9.0$, with $u(0)=5.0$, $\eta(0)=1.0$).}
\label{slug}
\end{center}
\end{figure}

In {\it Dictyostelium} a multicellular aggregate, the mound, is formed
by the cAMP-mediated aggregation of $10^2 - 10^5$ cells. This mound
elongates into a migrating cylindrical "slug", which under appropriate
conditions transforms into a mushroom-like structure, the fruiting
body, with a stalk composed of vacuolated, dead cells topped by a mass
of spores. Cell type pattern in the slug is highly organized along an
anterior-posterior axis (see Fig.~\ref{slug}). The anterior is
composed of prestalk (pst) cells (which ultimately differentiate into
stalk cells), with prestalk A cells (pstA) occupying the $\sim 10 \%$
foremost tip of the slug and prestalk O cells (pstO) $\sim 15 \%$
immediately following. The remaining $70-90 \%$, the posterior region,
is mainly made of prespore (psp) cells (which become spores) (see
review \cite{brn99}).

In slugs the qualitative cell type patterning is invariant over three
orders of magnitude of volume \cite{bnn57}. Furthermore slugs can
regenerate a normal cell-type pattern after removal of one of the cell
types \cite{rpr40}. These observations led Bonner to propose that the
proportion of cell types is regulated to have a constant value
\cite{bnn57}. Using an improved cell type marker, a
$\beta$-galactosidase with 1 hour half-life under the control of the
prestalk-specific promoter $ecmAO$, we have characterized the
regeneration process with previously unattained accuracy \cite{rfl00}
(see Fig.~\ref{slug}).

When we isolated the prestalk zones from large slugs, which originally
contained $10.5 \pm 3.3\%$ prestalk cells, the prestalk proportion at
the end of regeneration was $28.0 \pm 2.1\%$, significantly higher
than their initial value.  The initial proportion is thus not
completely recovered.  Moreover, removing a part of the prespore zone
fails to trigger any regeneration at all as long as the amputated slug
contains less than $\sim 30 \%$ pretalk cells; the regeneration
process is initiated, however, as soon as this threshold is exceeded.

From these results it follows the regulation is intrinsically
imprecise: it appears to control the proportion above and below the
$10\%/30\%$ thresholds, but not within the allowed range of values
that these thresholds define.

Several studies have presented evidence that prestalk cells require a
chemical secreted by prespore cells to remain in their prestalk
differentiated state \cite{iny89}. Based upon these facts, it has been
proposed that a small diffusible molecule acting as prespore inhibitor
might be regulating the proportion of cell types
\cite{bls86,lms93,nnj95}.

In contrast to previous suggestions \cite{mcw79}, the overall pattern
in the slug does not appear to be regulated by concentration
gradients, as in Turing-type reaction-diffusion mechanisms, but by
cell sorting. Thus (i) slugs don't seem to display any
size-independent characteristic length \cite{rfl00}; (ii) cell sorting
occurs faster than cell transdifferentiation; (iii) proportion
regulation without spatial pattern has been observed \cite{iny89}.
Sorting itself appears to be oriented by period cyclic AMP signals
propagating as waves from the tip \cite{drm00}. It therefore appears
to be justifiable to use models in which gradients are neglected and
the concentration of the prespore inhibitor is assumed to be
homogeneous along the slug.

The model we develop here is based on Kay's \cite{kay99}. We assume a
global negative feedback meant to regulate the proportion. This
feedback is mediated by a diffusible prespore inhibitor $u$ which is
produced by prespore cells at a rate $k_u$ and degraded by prestalk
cells at a rate $h_u(u)$. $n_{psp}$ and $n_{pst}$ are prespore and
prestalk cell density, respectively. Cell type transdifferentiation is
represented by the decreasing function $f(u)$ for $pst \Rightarrow
psp$ conversion and the increasing function $g(u)$ for $psp
\Rightarrow pst$.

\begin{equation}
\frac{du}{dt} = k_u n_{psp} - h_u(u) n_{pst}
\end{equation}

\begin{equation}
\frac{dn_{pst}}{dt} = g(u) n_{psp} - f(u) n_{pst}
\end{equation}

The degradation rate of $u$ is approximated as $h_u(u)=h_u u$. Cell
densities can be written as $n_{psp}=\rho (1-\eta)$ and $n_{pst}=\rho
\eta$, where $\rho$ stands for the total cell density and $\eta$ for
the prestalk proportion. Equations can be nondimensionalized with the
following change of variables: $t^*=\tau_u t$, $u^*=\frac{h_u}{k_u}u$,
$f^*(u^*)=\tau_{\eta}^{-1} f(u)$, $g^*(u^*)=\tau_{\eta}^{-1} g(u)$.
$\tau =\tau_{\eta} / \tau_u$, where $\tau_u= h_u \rho$ is the
characteristic time of $u$ degradation and $\tau_{\eta}$ the
characteristic time of cell conversion.  In the following we use the
new variables without the asterisk.

\begin{equation}
\frac{du}{dt} = (1-\eta) - u \eta
\end{equation}

\begin{equation}
\tau \frac{d \eta}{dt} = g(u) (1-\eta) - f(u) \eta
\label{eq-trans}
\end{equation}

\begin{figure}[htbp]
\begin{center}
  \includegraphics[width=15cm]{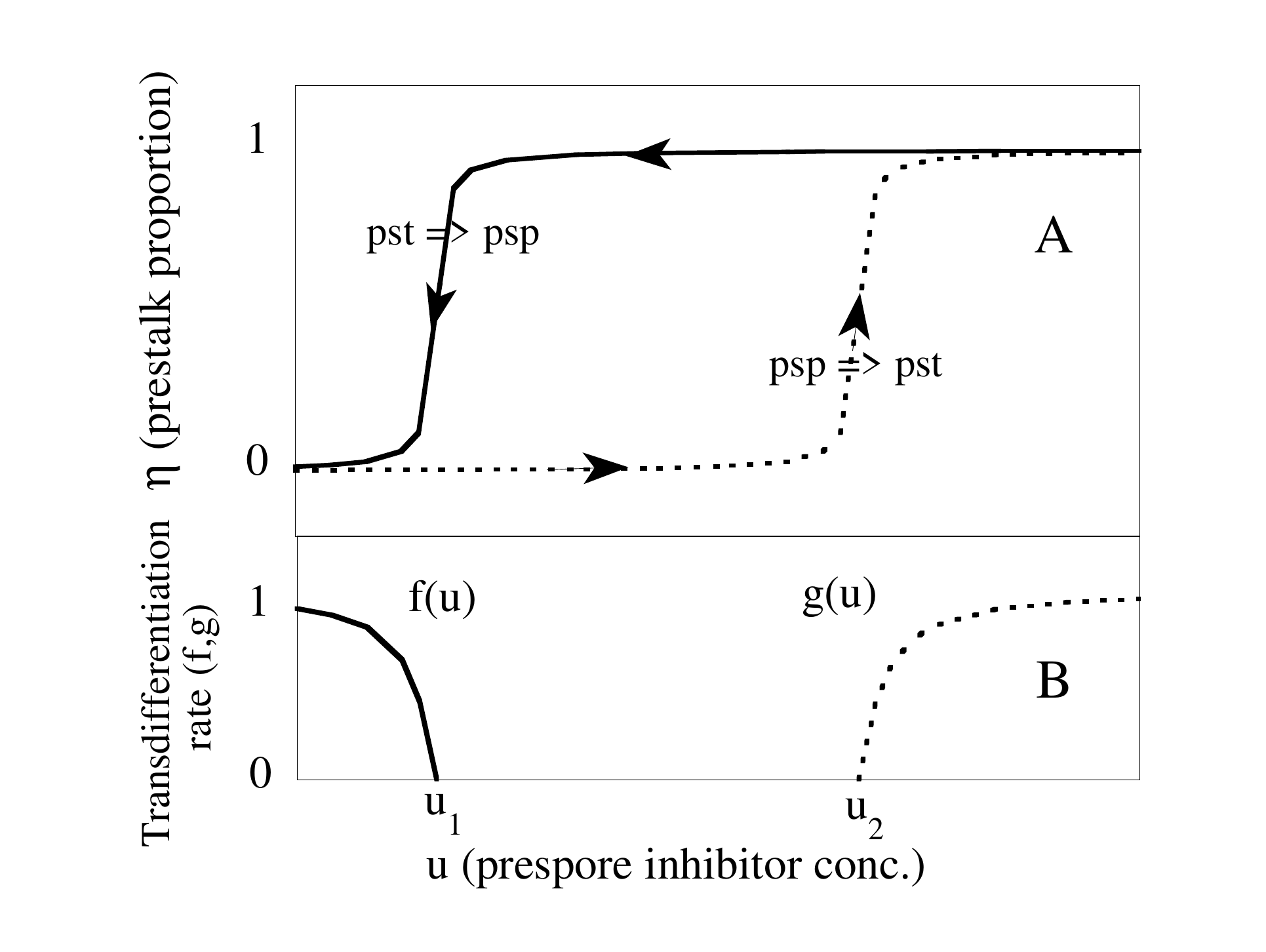}
\caption{(A) Prestalk proportion dependence on the prespore
  inhibitor/prestalk inducer. The hysteresis-like behaviour can be
  seen as the fingerprint of bistability in cell type differentiation
  (B) Cell type transdifferentiation rates $f(u)$ and $g(u)$. It is
  postulated that $f$ and $g$ only take a positive value below (above)
  some threshold $u_1$ ($u_2$).}
\label{hyste}
\end{center}
\end{figure}

The main novelty of this model is the hysteresis-like behaviour
introduced by the assumptions on $g$ and $f$, representing a
cell-autonomous positive feedback in the differentiation process.
First we assume that differentiation is bistable, i.e. $psp
\Rightarrow pst$ transdifferentiation begins only at $u>u_2$, but the
reverse conversion requires a decrease of $u$ below a much lower
threshold $u_1$. Second, as illustrated by Fig.~\ref{hyste}B, we
assume that this behaviour applies not only for the single cell, but
for the whole population (this can be shown to be the case as far as
the distribution of thresholds $u_1$ and $u_2$ don't overlap). Then,
if $u$ is taken as an external parameter, it follows from the
equation~\ref{eq-trans} that the proportion will display a
hysteresis-like behaviour such as shown in Fig.~\ref{hyste}A. Note
that if $f$ and $g$ had been allowed to overlap, the steady state
would be achieved at the expense of a continuous to-and-fro cell
interconversion (not observed experimentally) and $\eta$ vs. $u$
wouldn't display hysteresis \cite{nnj95}.

After the nondimensionalization, the equations have only 3 parameters:
$\tau$, $u_1$ and $u_2$. The particular dependence of $f$ and $g$ on
$u$ doesn't appear to have any noteworthy effect on the regulation
dynamics. Simulations have been performed assuming $f(u)=u_1-u$ for
$u<u_1$, otherwise zero, and $g(u)=u-u_2$ for $u>u_2$, otherwise zero.

\begin{figure}[htbp]
\begin{center}
  \includegraphics[width=15cm]{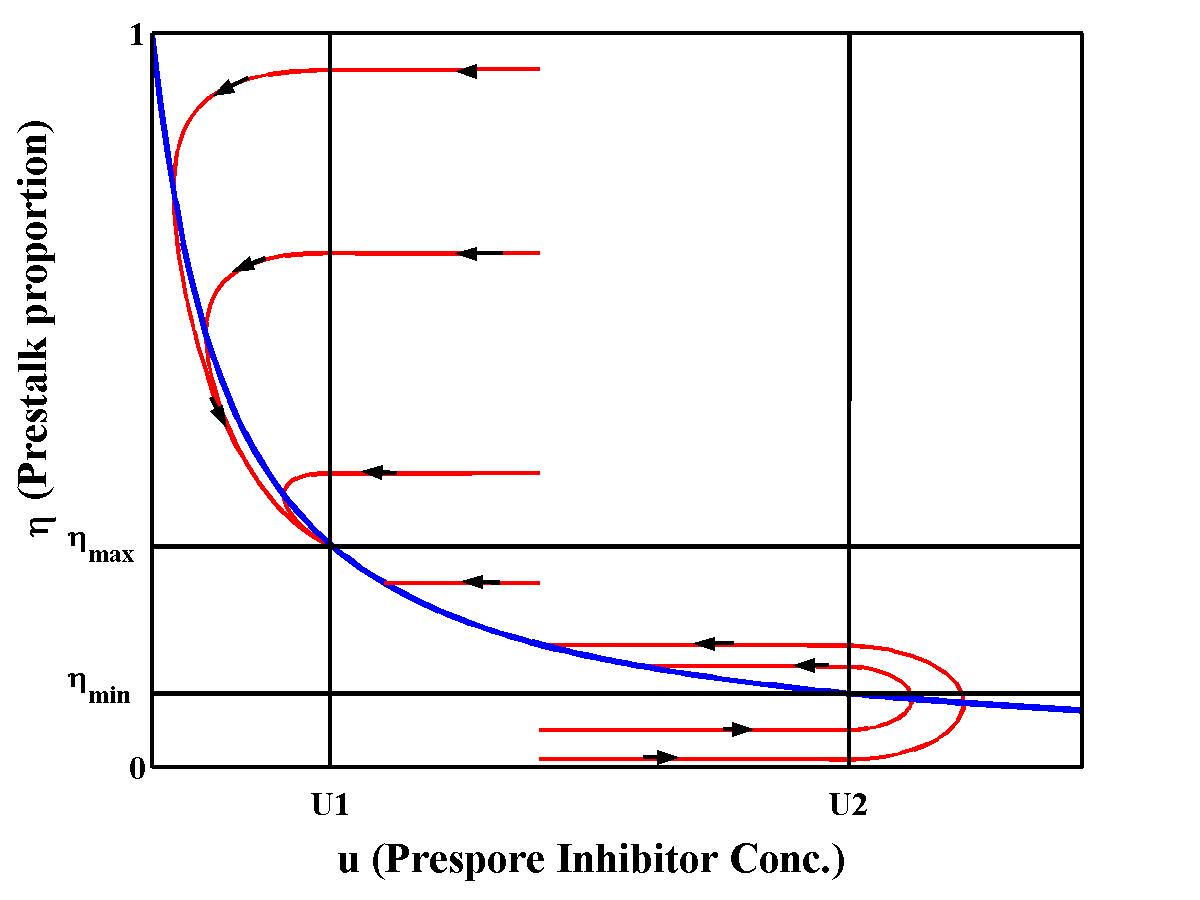}
\caption{(A) Trajectories obtained from simulations starting at
  different initial conditions for $\tau=30$, $u_1=2.3$, $u_2=9$ (red
  lines) . The nullcline $\frac{du}{dt}=0$ is shown in blue.  For
  initial proportions in the $\eta_{min} \le \eta \le \eta_{max}$,
  proportion doesn't change. However, when the system is perturbed
  beyond the upper/lower thresholds, the concentration of the
  regulatory molecule $u$ decreases/increases until it induces
  transdifferentiation and the proportion returns to the allowed
  range.}
\label{nullclines}
\end{center}
\end{figure}

Fig.~\ref{nullclines} shows the phase plane of the system.  A whole
segment of fixed points is found for $u_1 < u < u_2$ and $\eta =
\frac{1}{1+u}$. In consequence the range of stable proportions extends
from $\eta_{min}= \frac{1}{1+u_2}$ to $\eta_{max} = \frac{1}{1+u_1}$.
The experimental values of prestalk proportion in the slug ($10-30\%$
\cite{rfl00}) are satisfied for $u_1=2.4$ and $u_2=9$.

Perturbations in the proportion within the $(\eta_{min}, \eta_{max})$
interval result in an adjustment of $u$, but don't require the
proportion to be recovered, in accordance to recent findings
\cite{rfl00}. However, if the proportions are altered beyond
$\eta_{min}$ or $\eta_{max})$, proportion regeneration occurs after a
time interval of $u$ degradation/production $\tau_u$ (horizontal lines
in phase plane) and a time interval of exponential-like
increase/decrease of the proportion $\tau_{\eta}$ (vertical curves).
The parameter $\tau = \tau_{\eta}/\tau_u$ controls the ratio between
these caracteristic times. Simulations with any $\tau>30$ give an
excellent fit to experimental data of regeneration (see
Fig~\ref{slug}B), suggesting that $\tau_u \ll \tau_{\eta}$. Such a
result is fully consistent with observations that degradation time
$\tau_u$ of putative regulators might lie in the $10^{-1}-10^{-3}$
hours range, while the cell type transdifferentiation time
$\tau_{\eta}$ can be estimated in the 1-5 hours range.

One might expect that after an increase/decrease in the proportion,
the regulation would bring the proportion exactly to the lower/upper
thresholds of the available range $(\eta_{min}, \eta_{max})$. However
regeneration dynamics tends to overshoot beyond the thresholds. This
overshooting might eventually lead to damped oscillations in prestalk
proportion during regeneration. Being that such oscillations are both
unreallistic and undesirable, they may give us a broad hint of the
parameter ranges that natural selection would avoid. This is: small
$\tau$ (i.e. similar time scale in signalling and cell type
conversion) and narrow ranges of $(u_1$, $u_2)$ (i.e. short distance
between forward and reverse thresholds of $u$ for conversion).
Interestingly, this means that (at least some of) the parameters that
would produce a quick and precise regulation of the proportion, would
have a destabilising effect on the regulation dynamics. Current
estimates of the parameters lie far from this oscillating regime.

In spite of the success of the model in reproducing the dynamics of
proportion regeneration, a number of questions deserve further
inquiry:

(a) The actual molecular basis of the negative feedback is still yet
controversial. DIF-1 (Differentiation Inducing Factor-1), a
chlorinated alkyl hexaphenone, is the main candidate molecule for the
postulated prespore inhibitor/prestalk inducer \cite{kay99}. However
it has been recently found that it is necessary for pstO induction but
not for pstA \cite{thm00}.

(b) The observation that $\eta$ decreases with slug size
\cite{rfl00,nnj95} can be related to the decrease of $\mathrm{O_2}$
average concentration with size, which in turn may decrease production
rate $k_u$.

(c) We have assumed that the $u$ concentration is homogeneous along
the slug. However, the diffusion length $\lambda_u$ of putative
regulatory molecules such as DIF-1 or cAMP can be estimated in the
$0.15-1.5$ mm range. Since effective diffusion will be even smaller,
gradients may play a role in medium and big ($\sim 1.0$ mm long)
slugs. Being that the prespore inhibitor gradient is reversed, the
difference in the transdifferentiation thresholds $u_1$ and $u_2$
becomes essential to stabilize the pattern.

(d) The transdifferentiation rates $f$ and $g$ for the population
dynamics have been assumed based on the observation of hysteresis in
cell type conversion. They should eventually be derived from positive
feedback loops in the signal transduction pathways at the single cell
level \cite{brn99}.

(e) This study has been solely concerned on the regulation between
already differentiated cell types. It should be stressed that it may
not be immediately extensible to the initial {\it Dictyostelium} cell
type differentiation where positional effects may be important
\cite{erl95}.

A general lesson may be drawn from this study. In tissue maintenance,
it is important both to control the cell type proportions and to keep
each of these cell types in a well differentiated state. However,
there seems to be a conflict of between these two requirements. On the
one hand proportion regulation is better served by signalling that
involves fast global negative feedback. On the other hand, a robust
cell differentiation requires a strong cell-autonomous positive
feedback to operate the ``switch'' between cell types. Without
positive feedback cell differentiation would result in a continuous
spectrum of cell phenotypes. Yet, as a result of the positive
feedback, cell differentiation will always display some hysteresis in
respect to the control exerted by the global regulative mechanism.
This hysteresis poses a limit to the precision of the proportion
regulation. In other words, it looks as though the more robust is the
cell differentiation the less precise is the proportion regulation and
vise versa. This conflict between precision and bistability appears to
be a general property of systems of globally coupled bistable elements
\cite{mzg95}.

In spite of its sheer simplicity, we believe that the model proposed
stands on reasonable experimental foundations and delivers new
conceptual insights. In particular, it provides an explanation for
recent findings on the imprecision of {\it Dictyostelium} cell type
proportioning. In general, it illustrates the possible conflict
between proportion regulation and robustness of cell differentiation
in multicellular tissues.

IR would like to thank K. Inouye for insightful comments and sharing
of unpublished data. He is also grateful to A. Amagai for continuous
support and to S. Sawai, M. Sano and T. Mizuguchi for fruitful
discussions.

\end{document}